\documentclass[aps,prx,reprint,groupedaddress]{revtex4-1}

\usepackage[modulo]{lineno}

\usepackage{amsmath,amsfonts,
	amsgen,amsbsy,amsopn,amstext,amssymb
}
\usepackage{booktabs}
\usepackage{graphicx}
\usepackage{dcolumn}
\newcolumntype{.}{D{.}{.}{-1}}
\usepackage{epstopdf}
\usepackage{color}

\usepackage[
colorlinks=true,
linkcolor=blue,
citecolor=blue,
filecolor=blue,
urlcolor=blue]{hyperref}

\begin{document}
	
	
	\title{Holograms to focus arbitrary ultrasonic fields through the skull}
	
	\author{Sergio~Jim\'enez-Gamb\'in}
	\author{No\'e~Jim\'enez}
	\email{nojigon@upv.es}
	\author{Jos\'e~Mar\'ia~Benlloch}
	\author{Francisco~Camarena}
	\affiliation{Instituto de Instrumentaci\'on para Imagen Molecular, Consejo Superior de Investigaciones Cient\'ificas, Universitat Polit\`ecnica de Val\`encia. Camino de vera s/n 46022 Val\`encia, Spain} 
	
	\date{\today}
	
	\begin{abstract}
		We report 3D-printed acoustic holographic lenses for the formation of ultrasonic fields of complex spatial distribution inside the skull. Using holographic lenses, we experimentally, numerically and theoretically produce acoustic beams whose spatial distribution matches target structures of the central nervous system. In particular, we produce three types of targets of increasing complexity. First, a set of points are selected at the center of both right and left human hippocampi. Experiments using a skull phantom and 3D printed acoustic holographic lenses show that the corresponding bifocal lens simultaneously focuses acoustic energy at the target foci, with good agreement between theory and simulations. Second, an arbitrary curve is set as the target inside the skull phantom. Using time-reversal methods the holographic beam bends following the target path, in a similar way as self-bending beams do in free space. Finally, the right human hippocampus is selected as a target volume. The focus of the corresponding holographic lens overlaps with the target volume in excellent agreement between theory in free-media, and experiments and simulations including the skull phantom. The precise control of focused ultrasound into the central nervous system is mainly limited due to the strong phase aberrations produced by refraction and attenuation of the skull. Using the present method, the ultrasonic beam can be focused not only at a single point but overlapping one or various target structures simultaneously using low-cost 3D-printed acoustic holographic lens. The results open new paths to spread incoming biomedical ultrasound applications including blood-brain barrier opening or neuromodulation.
	\end{abstract}

\flushbottom
\maketitle

\thispagestyle{empty}

\section{Introduction}\label{sec:intro}
Holographic plates are surfaces that when illuminated by a wave, typically light, modify the phase of the transmitted or reflected wavefront in such a manner that a complex image can be formed \cite{gabor1948new,gabor1949microscopy,leith1962reconstructed}. 
In recent years, subwavelength thickness holographic metasurfaces have been designed using structured materials with subwavelength resonances, namely metamaterials \cite{ni2013metasurface,huang2013three}. Analogously, in acoustics, a broad range of locally-resonant structures have been proposed to obtain a precise control of the wavefront at a subwavelength scale \cite{ma2016acoustic,cummer2016controlling}, including effective negative mass density \cite{liu2000locally} and/or bulk modulus metamaterials \cite{fang2006ultrasonic,yang2013coupled}. Acoustic metamaterials allow an accurate control of the reflected \cite{li2013reflected,xie2014wavefront,jimenez2017metadiffusers,jimenez2017rainbow, qi2017acoustic} or transmitted wavefronts \cite{bok2018metasurface,li2015metascreen,li2015three}. The use of these structures has been exploited to design negative-refraction superlenses \cite{kaina2015negative} or hyperbolic dispersion-relation hyperlenses \cite{li2009experimental} that exhibit subwavelength focusing properties in the near field. 
Holographic lenses have also been reported in acoustics to generate complex acoustic fields \cite{melde2016holograms,xie2016acoustic,zhu2018fine,memoli2017metamaterial}. Multi-frequency holograms have been also reported \cite{brown2017design}. Equivalently, using phased-array sources it has been reported the generation of complex beam patterns \cite{hertzberg2011bypassing}, self-bending and bottle beams\cite{zhang2014generation}, or vortex beams for particle levitation and manipulation \cite{marzo2015holographic}. Mixed approaches between metamaterials and phased arrays have been also presented \cite{norasikin2018soundbender}.  

In these applications, holographic lenses have demonstrated the ability to manipulate acoustic waves in free media, i.e., without inhomogeneities. However, when using ultrasound in biomedical engineering applications, acoustic beams encounter in their path multiple tissue layers of complex geometry with non-homogeneous properties. For instance, an accurate control of the focused beam is at the basis of focused ultrasound therapy techniques, e.g., as in high intensity focused ultrasound hyperthermia, thermal ablation or histotripsy, or in extracorporeal shockwave lithotripsy \cite{ter2007high,bouakaz2016therapeutic}. Focusing directly into human soft-tissues can efficiently be achieved by using conventional systems as ultrasound beam aberrations are typically small in these media \cite{szabo2004diagnostic}. However, when the target tissue lays behind high-impedance tissues, e.g., soft-tissue surrounded by bones, the beam experiences strong aberrations due to refraction, reflection and absorption processes \cite{gelat2014comparison}. Some applications make use of existing acoustic windows by targeting tissues from specific locations. Nevertheless, in the case of transcranial propagation skull bones are always present in the path towards the central nervous system (CNS). In this way, the precise control of acoustic focus into the CNS is mainly limited due to the strong phase aberrations produced by the refraction and attenuation of the skull \cite{fry1978acoustical}.

To overcome these limitations, minimally-invasive techniques were developed in the past to design active focusing systems using the time-reversal invariance of the acoustic propagation \cite{thomas1996ultrasonic} or phase conjugation methods \cite{hynynen1998demonstration}. In minimally-invasive techniques, a small acoustic source is introduced into the skull, together with a biopsy catheter. When the catheter reaches the target tissue it radiates a short ultrasonic pulse that travels outwards and it is recorded by a hemispherical multi-element array surrounding the patient's head. Then, the elements of the phased-array are set to re-emit the time-reversed recorded waveforms (or phase conjugated harmonic signals). Due to spatial reciprocity and time-reversal invariance of the acoustic media, the generated wavefront focuses at the catheter location, i.e., at target tissue \cite{thomas1996ultrasonic}. Later, it was demonstrated that non-invasive versions of these techniques can be obtained using numerical simulations \cite{sun1998focusing,aubry2003experimental}. In these techniques a tomographic image is previously obtained from patient's head to extract the geometry of the skull and its acoustic properties \cite{aubry2003experimental}. Using full-wave simulations the time-reversed wavefront is calculated by exciting the simulation with a virtual source at the desired focal spot. Then, a physical hemispherical phased-array is excited with the synthetic time-reversed waveforms \cite{tanter1998focusing} or phase conjugated signals \cite{sun1998focusing}, and sharp focusing through skull-aberration layers is retrieved. Other techniques include the optimization of phase-arrays using magnetic resonance imaging (MRI) to maximize acoustic-radiation-force induced displacements into the target focus \cite{hertzberg2010ultrasound}. However, up-to-date phased-array systems are restricted to a limited number of channels, e.g., 1024 for the Exablate{\small\textsuperscript{\textregistered}} Model 4000 (InSightec, Ltd)\cite{jolesz2014intraoperative}, that can be insufficient to fully record the required holographic information in order to conform a complex beam pattern. In addition, phased-arrays are effective, but due to its high economical cost it is desirable to use passive structures to control acoustic beams.

\begin{figure}[t!]
	\centering
	\includegraphics[width=0.9\linewidth]{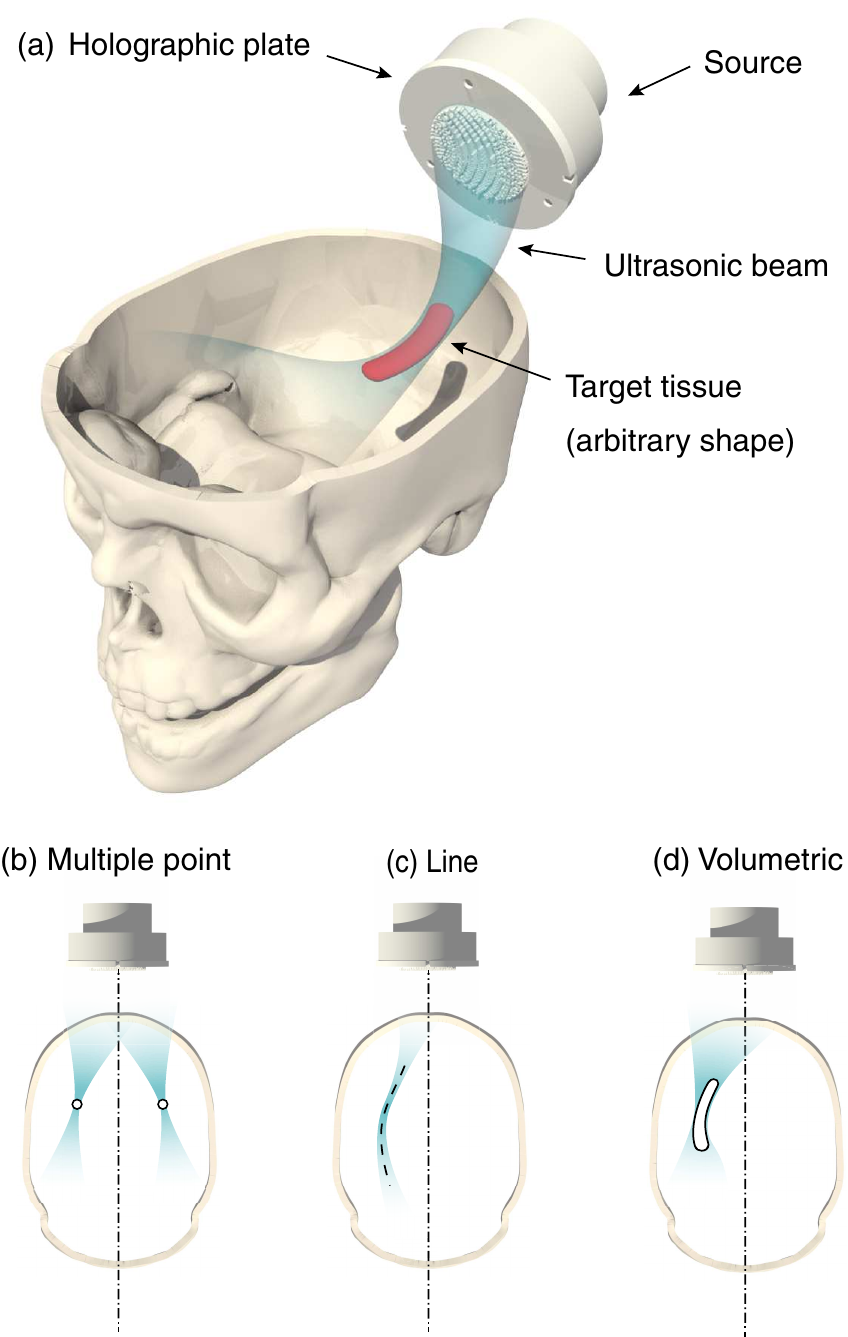}
	\caption{(a) Scheme of the holographic lens focusing over a target CNS structure. (b) Focusing on a set of arbitrary points (bifocal holographic lens), (c) Focusing over arbitrary line (self-bending holographic lens), (d) Focusing over an arbitrary volume (volumetric holographic lens).}
	\label{fig:scheme}
\end{figure}

Only few theoretical works have tackled the problem of beam focusing through aberrating layers using metamaterials \cite{shen2014anisotropic} or phase plates \cite{maimbourg20183d,FERRI2019867}. In Ref. \cite{shen2014anisotropic} a 2D configuration was proposed theoretically using a metasurface based on membranes. Recently, the use of phase plates to generate simple focused sources have been reported to avoid beam aberrations in transcranial propagation \cite{maimbourg20183d}. However, the technique was limited to focus the beam into a single focal spot at the near field of the source. Besides, in some non-thermal transcranial ultrasound applications such as blood-brain barrier opening \cite{hynynen2001noninvasive} or neuromodulation \cite{tyler2008remote} the ultrasound beam might be set to fully-cover a geometrically complex CNS structure rather than focusing over a small focal spot.

In this work, we propose the use of 3D-printed holographic phase plates to produce ultrasonic fields of arbitrary shape into the human brain. 
The holographic lenses designed in this work allow the reconstruction of complex diffraction-limited acoustic images including the compensation of the aberrations produced by a skull phantom. 

In particular, we theoretically, numerically and experimentally demonstrate the generation of several holographic patterns, of increasing complexity, all with direct practical application to biomedical ultrasound: an arbitrary set of points, an arbitrary curved line, and an arbitrary volume. First, we provide the conditions to generate a simple holographic pattern, i.e., a set of diffraction-limited focal points, as sketched in Fig.~\ref{fig:scheme}~(b). 
In particular, we extend the use of holographic lenses to generate bifocal beams, matching both foci simultaneously the location of left and right human hippocampus. Second, we demonstrate that ultrasonic beams with curved trajectory along the internal CNS tissues can also be produced, as Fig.~\ref{fig:scheme}~(c) shows. In this way, the acoustic beam can be bent following arbitrary paths producing a self-bending beam inside the CNS. Finally, we report the generation of a beam pattern that overlaps with the volume of a specific CNS structure, as shown in Fig.~\ref{fig:scheme}~(d), in particular we target the right human hippocampus.

\begin{figure*}[t]
	\centering
	\includegraphics[width=1\textwidth]{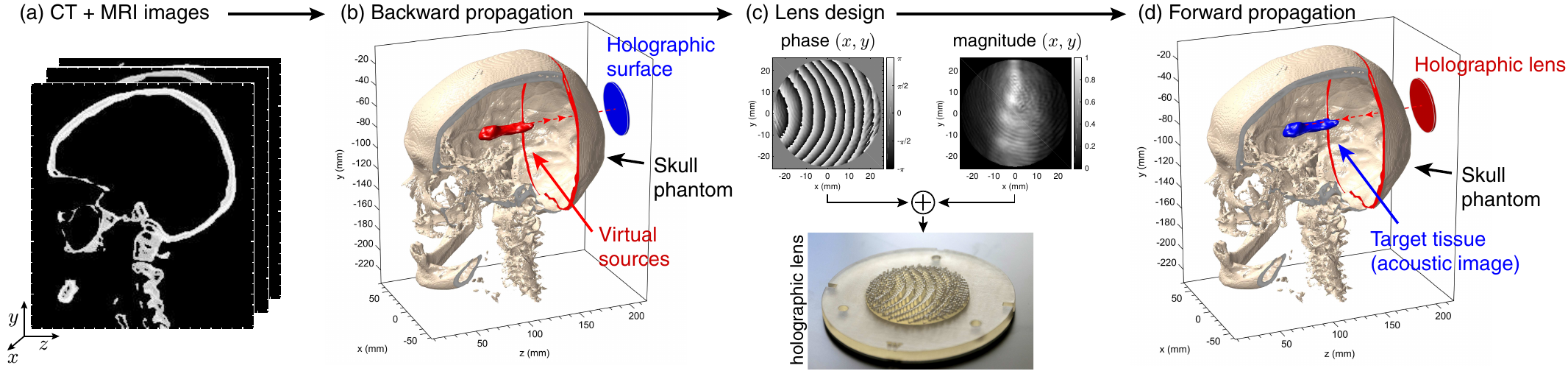}
	\caption{Hologram generation process. (a) CT+MRI tomographic images. (b) Selected target (red volume) acting as a virtual acoustic source and holographic recording surface (blue area), (c) Lens design using the TR back-propagated field. (d) Forward propagation from the holographic lens (red area) to the target tissue (blue volume).}
	\label{fig:methods}
\end{figure*}

\section{Methods}
The process of hologram generation is composed of four steps. First, we extract the geometry and acoustic properties of a human skull from X-ray CT images, as shown in Fig.~\ref{fig:methods}~(a), and from MRI tomographic images we identify the target tissue structure, e.g., the right human hippocampus as shown Fig.~\ref{fig:methods}~(b). Second, a back-propagation method is used to calculate the acoustic wavefront generated from a set of virtual sources and impinging on a holographic surface located outside the skull phantom, as shown in Fig.~\ref{fig:methods}~(b). Third, the phase-plate lens is generated by using the phase and amplitude of the recorded wavefront at the holographic surface, as shown in Fig.~\ref{fig:methods}~(c). Finally, the lens is excited with a flat and uniform ultrasonic transducer and the target acoustic image is reconstructed by either theoretical, numerical forward-propagation or experimental methods, as shown in Fig.~\ref{fig:methods}~(d).

\subsection{Tomographic image acquisition}
First, in order to model the skull geometry, we used the CT Datasets of a female human head with an isotropic resolution of 1 mm (interpolated to 0.22 mm for the numerical simulation) from the National Library of Medicine's Visible Human Project available for general use by the University of Iowa. Experiments were conducted in a 3D printed skull phantom, while, in addition, we included full-wave simulations using the acoustical properties of the skull bones. Thus, for the skull phantom simulations we used homogeneous acoustical parameters matching those of the 3D printing material, while for the realistic skull simulations we used the same geometry but the inhomogeneous acoustical parameters of the skull were derived using the same the CT data, converting the apparent density tomographic data in Hounsfield units to density and sound speed distributions using the linear-piecewise polynomials proposed in Refs.~\cite{schneider1996calibration,mast2000empirical}. 

After, we used a human atlas made publicly available by the International Consortium for Brain Mapping (ICBM) from the Laboratory of Neuro Imaging \cite{mazziotta1995probabilistic}. This atlas provided us T1 weighted  MRI data that was used to identify the shape and location of the human hippocampus. In particular we used for segmentation the ITK-SNAP software \cite{yushkevich2006user} to obtain the shape and location of the left and right hippocampi.

\subsection{Calculation methods}
We use two methods, of increasing complexity, to estimate the back-and-forward acoustic fields: a semi-analytical method using Rayleigh-Sommerfeld diffraction integral and a pseudo-spectral time-domain simulation method.

On the one hand, for theoretical calculations in homogeneous media, i.e, in water without the skull phantom, the acoustic pressure field given by $p({{\bf r}})$ at point ${\bf r}$, generated by a moving surface $S$ of arbitrary shape located at coordinates ${\bf r}_0$ and vibrating with a complex particle velocity $v_0({\bf r}_0)$ normal to the surface, is given by the Rayleigh-Sommerfeld diffraction integral \cite{Blackstock2000}:
\begin{align}\label{eq:ray}
p({\bf r},\omega) = \frac{i\omega \rho_0}{2\pi}\int_{S}\frac{v_0({\bf r}_0)\exp\left(-k_0\left|{\bf r} - {\bf r}_0\right|\right)}{\left|{\bf r} - {\bf r}_0\right|} dS,
\end{align}
where $\omega=2\pi f$; $k_0=\omega/c_0$, $c_0$ and $\rho_0$ are the wavenumber, sound speed and density of water. Note that in Eq.~(\ref{eq:ray}) diffraction is captured exactly as compared with angular spectrum methods, so it can be applied to high-aperture sources. 

On the other hand, for calculations including aberration layers we use a pseudo-spectral simulation method with $k$-space dispersion correction to numerically integrate the linearized constitutive relations of acoustics \cite{treeby2010modeling,treeby2012modeling}. In an inhomogeneous and absorbing media, the governing equations, i.e., the continuity equation, the momentum conservation equation and the pressure-density relation, can be written as three-coupled first-order partial differential equations as:
	\begin{align}
	\frac{\partial \rho}{\partial t} &= -\rho_0\nabla \cdot {\bf u} - {\bf u} \cdot \nabla\rho_0, \label{eq:kspace_mass}\\
	\frac{\partial \bf u}{\partial t} &= -\frac{1}{\rho_0}\nabla p ,\label{eq:kspace_momentum}\\
	p &= c^2_0\left(\rho + {\bf d} \cdot\nabla\rho_0 - \mathrm{L}\rho\right), \label{eq:kspace_pressure}
	\end{align}
	where ${\bf u}$ is the acoustic particle velocity, ${\bf d}$ is the acoustic particle displacement, $p$ is the acoustic pressure, $\rho$ is the acoustic density, $\rho_0$ is the ambient (or equilibrium) density, $c_0$ is the sound speed, and $L$ is a linear operator introducing the frequency-dependent absorption and dispersion \cite{treeby2010modeling}. Tissue absorption following a power-law on frequency given by $\alpha (\omega) =\alpha_0\omega^\gamma$, where $\alpha_0$ is the absorption coefficient and $\gamma$ is the exponent of the frequency power law, together with its corresponding physical dispersion are included by the integro-differential operator as:
	\begin{align}
	\mathrm{L} = \tau \frac{\partial }{\partial t} \left(-\nabla^2\right)^{\frac{\gamma}{2}-1} + \eta\left(-\nabla^2\right)^{\frac{\gamma+1}{2}-1},
	\end{align}
	where $\tau = -2\alpha_0 c_0 ^{\gamma-1}$ and $\eta = 2\alpha_0 c_0^\gamma\tan\left(\pi \gamma/2\right)$ and the absorption and dispersion proportionality coefficients. This operator is solved efficiently using the fractional Laplacian in the $k$-space. This simulation method was selected as it provides low numerical dispersion as compared with finite-differences methods \cite{jimenez2016time}. We used a numerical grid with a spatial step of $\Delta x=\Delta y = \Delta z = 223 ~\mu$m and a numerical temporal step of $\Delta t = 19.1$ ns, leading to a Courant-Friedrichs-Lewy number \cite{treeby2010modeling} of 0.13 in water and an spatial sampling of 6 grid points per wavelength in water for a frequency of 1 MHz. These parameters were fixed to all simulations in this paper. 

\begin{figure}[t]
	\centering
	\includegraphics[width=1\columnwidth]{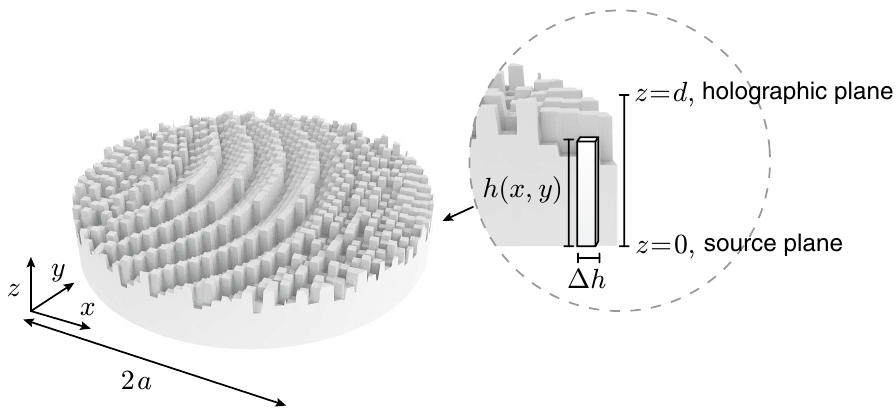}
	\caption{{Geometry of the holographic lens. The lens, of aperture $2a$ is subdivided in pixels of height $h(x,y)$. The source is located at $z=0$, while the holographic plane is located at $z=d$.}}
	\label{fig:lens}
\end{figure}

\subsection{Lens design}
{Under the assumption of reciprocity, time-invariance and linearity of the system, a time-reversal (TR) technique together with a direct method was used to design the only-phase holographic lens.}

First, we set some virtual sources inside the skull phantom and the back-propagated field was estimated at a given surface outside the skull phantom. For the bifocal lens, two virtual sources were set as monopoles with same phase and amplitude, located at the center of mass of the two hippocampi (right and left), {as sketched in Fig.~\ref{fig:scheme}~(a)}. For the self-bending beam, a set of 50 virtual sources were located following an arbitrary curve {as sketched in Fig.~\ref{fig:scheme}~(b)}, each source compensated by a phase factor of $\exp(ik_z z)$ accounting for the direction of arrival of the wavefront. Finally, for the volumetric hologram, as sketched in Fig.~\ref{fig:scheme}~(c), a set of virtual sources were spatially distributed with a separation of $\lambda/6$ ({to match the numerical grid used}) over a sagittal plane of the right human hippocampus. The recorded field was captured at a given surface, i.e, at a holographic surface, outside the skull phantom.

Second, the recorded conjugated pressure distribution at the working frequency was used to design the physical lens. 
The lens surface was divided in squared pixels of different height, $h(x,y)$ and uniform width, $\Delta h$, {as shown in Fig.~\ref{fig:lens}.} We assume each elastic column to vibrate longitudinally as a Fabry-P\'erot resonator. For each column, the field at the holographic plane located at ${\bf x}_0=(x,y,d)$ is given by the complex transmission coefficient\cite{jimenez2017quasiperfect}:
\begin{align}\label{eq:fabryperot}
T({\bf x}_0) = \frac{2Z \mathrm{e}^{-i k_0 [d-h({\bf x}_0)]}}{2Z\cos\left[k_L h({\bf x}_0)\right]+i\left(Z^2+1\right)\sin\left[ k_L h({\bf x}_0)\right]},
\end{align}
where $d$ is the distance from the bottom of the lens ($z=0$) to the holographic surface, the normalized impedance is given by $Z = Z_L/Z_0$, and $Z_0=\rho_0 c_0$ is the impedance of water and $Z_L=\rho_L c_L$, $k_L=\omega/c_L$, $\rho_L$ and $c_L$, are the impedance, wavenumber, density and sound speed of the lens material. 
In order to obtain the height of each pixel of the lens, an analytic inversion of Eq.~(\ref{eq:fabryperot}) is not possible due to the trigonometric terms. Instead, we first numerically evaluate the expression for a broad range of pixel heights ranging from a {minimum height that was set to $h_{min}=5$ mm to guarantee structural consistency}, to a given height that provides a phase of the transmission coefficient 2$\pi$ greater than for $h_{min}$, i.e., $d=15$ mm, and using steps of 1 $\mu$m, well below the printer accuracy. Finally, we performed interpolation using a cubic-spline method to obtain the heigh of the pixel as a function of the required phase. In this way, by tuning the height of each Fabry-P\'erot resonator the phase at the output of each pixel can be tailored to that of a target holographic surface. 

However, using this kind of lenses the degree of freedom to modify the magnitude of the field at the holographic surface is limited. 
Iterative methods were employed in the past to obtain equivalent lenses only with phase distributions \cite{melde2016holograms}. In this work, iterative methods are prohibitive: the 3D simulations including aberration layers involve long calculation times, e.g., 20 hours in a Intel\textsuperscript{\textregistered} Xeon\textsuperscript{\textregistered} CPU E5-2680 v2 2.80GHz, 256 GB RAM, using a CPU parallel implementation of the code. 
Instead, we use a direct method to estimate an equivalent holographic lens of uniform field magnitude \cite{tsang2013novel}. The basis of this direct method is the sequential scanning of the pixels to modify the complex transmission coefficient. The method work as follows: First, the odd and even rows are scanned from opposite directions, and a bidirectional error of the diffusion process is calculated. The magnitude of each visited pixel is forced to be a constant value 
while the exact phase value is preserved. The resulting error is diffused to the neighboring pixels. Finally, the result gives a surface with a modified phase depending on the bidirectional error diffusion process \cite{tsang2013novel}. 
The main limitation of this method is that if the pixel width is small it will appear areas with isolated long pixels, i.e., columns, that can experience bending modes. Note this do not imply that a lens cannot be designed, but the theory presented here only apply to longitudinal modes on each pixel. The size of the pixels used in this work, $5/6$ times the wavelength, is thick enough to ensure that the resonance frequency of the first bending mode is far away from the first longitudinal Fabry-P\'erot resonance frequency.

\begin{figure*}[t]
	\centering
	\includegraphics[width=1\linewidth]{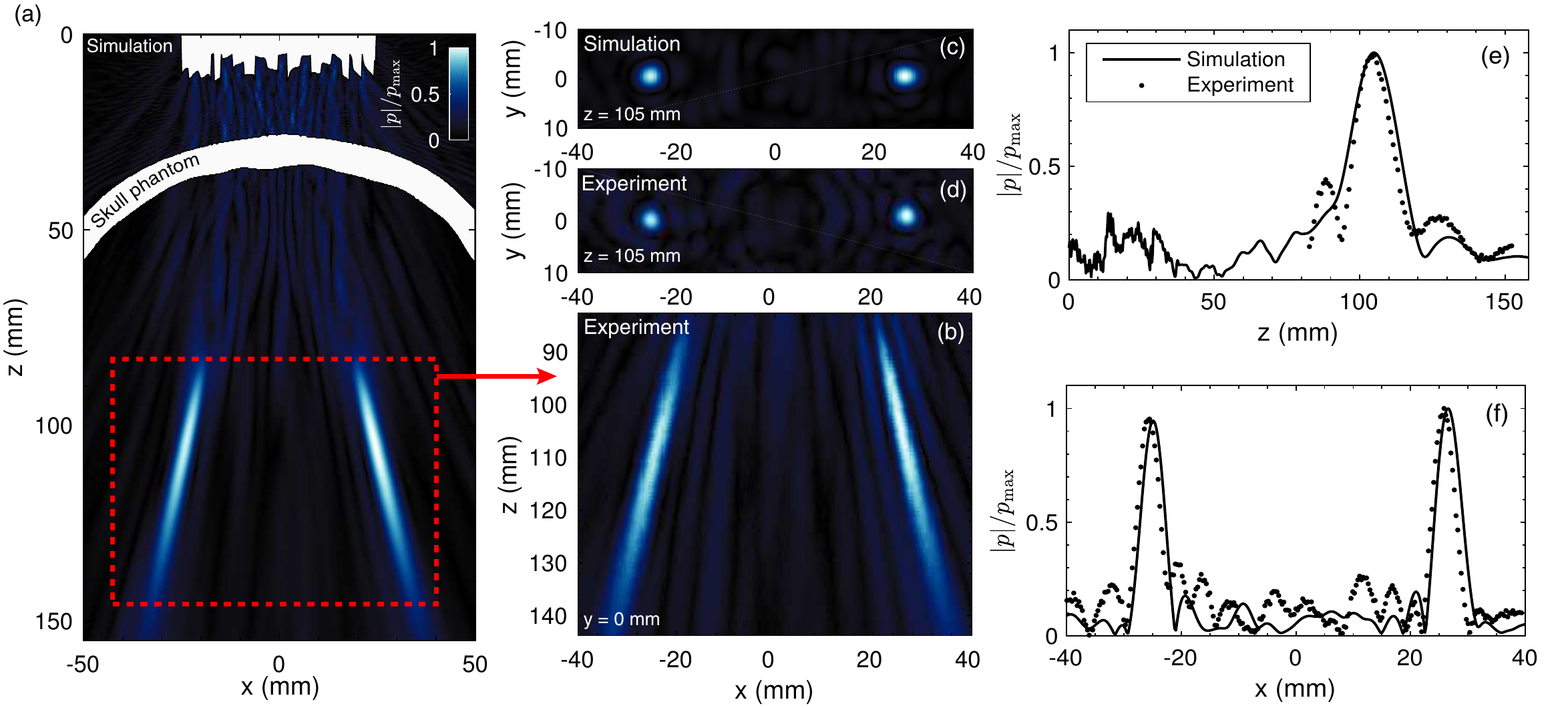}
	\caption{Axial field cross-section obtained for the bifocal lens using simulations (a) and experiments (b). (c-d) Corresponding transversal pressure field distributions. Colorbar in units normalized to the peak pressure. (e,f) Simulated and experimental normalized axial and transversal field cross-sections, respectively.}
	\label{fig:bifocal}
\end{figure*}

Third, lens with an aperture of $2a = 50$ mm were manufactured using 3D printing techniques. On the one hand, the bifocal holographic lens was manufactured by additive 3D printing techniques using Ultimaker 3 Extended (Ultimaker B.V., The Netherlands) with a resolution of 100 $\mu$m in both, lateral and axial directions and PLA material. As, in general, the height profile of the lens is smooth, we set the square pixel resolution to $\Delta h = 0.22$ mm. The acoustical properties of PLA material were obtained experimentally using a {pulse-echo technique} in a test cylinder, resulting in a measured sound speed of $c_L=1818$ m/s and a density of $\rho_L=1127$ kg/m\textsuperscript{3}, matching with those reported in existing literature \cite{lirette2017pla}, and the absorption was set to $\alpha=13.72$ dB/cm at 1.112 MHz \cite{lirette2017pla}. On the other hand, the self-bending and volumetric holographic lenses, which needed a more accurate printing technique for their complex pattern, were 3D printed using Polyjet techniques with an Objet30 printer (Stratasys, USA), with a resolution of 100 $\mu$m and 28 $\mu$m in lateral and axial directions respectively, and using a photo-resistive polymer (Veroclear\textsuperscript{\textregistered}, Stratasys, USA). As a result of the direct method to obtain the equivalent holographic lens of uniform field magnitude \cite{tsang2013novel}, the height distribution presents high spatial modulations (see Fig.~\ref{fig:methods}~(c)). Thus, the pixel resolution was increased to $\Delta h = 1$ mm to ensure each column vibrates as a longitudinal Fabry-P\'erot resonator avoiding bending modes around the working frequency for the volumetric hologram. For this material we experimentally estimated $c_L=2312$ m/s and $\rho_L=1191$ kg/m\textsuperscript{3} and $\alpha=3.06$ dB/cm at 1.112 MHz, matching the values reported in existing literature \cite{melde2016holograms}.

\subsection{Skull phantom}\label{sec:meth:skull}
The geometry of the skull phantom was extracted from the 3D CT images as described previously. The sound speed and density distributions were first estimated from the apparent density given by the CT images in Hounsfield units \cite{schneider1996calibration,mast2000empirical}. Then, as the 3D printing technique results in homogeneous material, the acoustic properties, including the absorption were considered uniform along the skull bone volume \cite{gatto2012three,robertson2017sensitivity,bai2018design}. The skull phantom was manufactured by additive 3D printing techniques using Ultimaker 3 Extended (Ultimaker B.V., The Netherlands) with resolution of 100 $\mu$m in both, lateral and axial directions and using PLA material. { The acoustic parameters for the 3D printed phantom are the same than for the PLA lenses.}

Finally, for the simulations using a realistic skull, the acoustical parameters were derived using the CT data, converting the apparent density in Hounsfield units to density and sound speed distributions using the linear-piecewise polynomials proposed in Refs.~\cite{schneider1996calibration,mast2000empirical}. The density data ranges between $\rho_{0} = 1000$ kg/m$^3$ (water) and $\rho_{max} = 2206$ kg/m$^3$ (bone), the sound speed values range between $c_{0}=1500$ m/s and $c_{max} = 3117$ m/s, matching those reported in literature \cite{hill2004physical,bouakaz2016therapeutic} and the bone absorption was set to 12.6 dB/cm at 1.112 MHz \cite{cobbold2006foundations}.

{The details about the measurement system can be found in Appendix \ref{appendix}. }


\section{Multiple-point holograms} 
We start with the bifocal holographic lens. First, two points located at the center of mass of both left and right human hippocampi are selected. Second, we set this pair of points as the location of virtual sources for the TR method. For the lens designs in free media, i.e., without the skull, we make use of the Rayleigh-Sommerfeld diffraction integral (see Methods section for further details). For the lens designs of holographic surfaces including the skull-aberration layers we make use of low-numerical-dispersion simulations based on pseudo-spectral methods \cite{treeby2010modeling}. In this way the simultaneous back-propagation of the fields irradiated by both virtual monopoles can be calculated at the holographic surface which is located at the rear part of the skull. The phase-plate lens is designed using the conjugated complex field recorded at the surface. Then, the lens is placed at the location of the holographic surface as shown in Fig.~\ref{fig:bifocal}~(a), and a forward-propagation calculation is carried out to test the quality of the reconstructed acoustic image.

\begin{figure*}[t]
	\centering
	\includegraphics[width=1\linewidth]{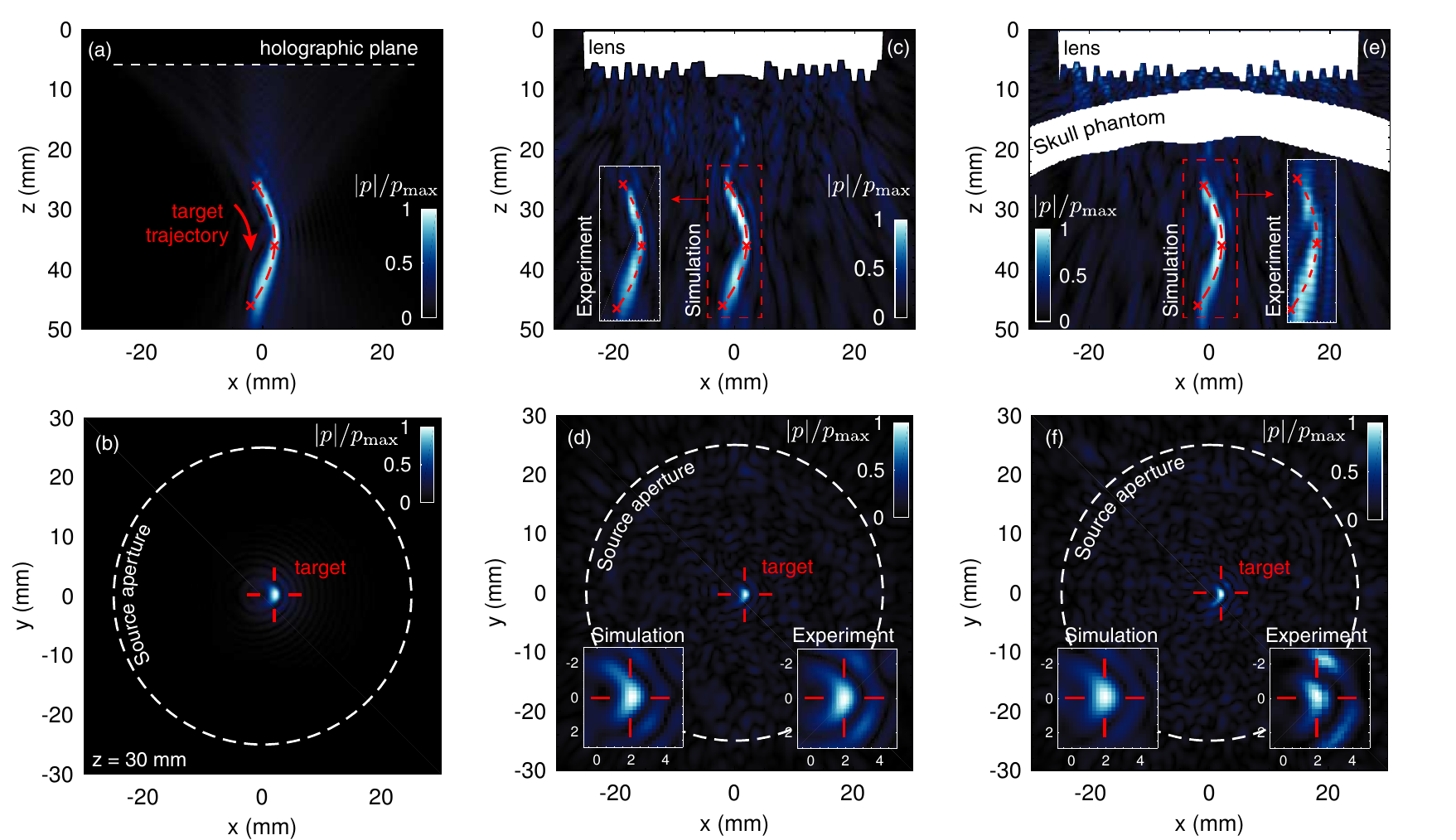}
	\caption{Theoretical (a) axial and (b) transverse pressure field distribution for the self-bending beam in water. Simulated (c) axial and (d) transverse pressure field distribution for the self-bending beam in water. Corresponding experimental results are shown in the insets in (c) and (d). (e,f) Simulated  axial and transversal pressure field including the skull phantom. Corresponding experimental results are shown in the insets in (e) and (f).}
	\label{fig:selfbending}
\end{figure*}

The field produced by the bifocal lens propagating through a human occipital/parietal skull phantom including the compensation for the aberrations of the skull is shown in Figs.~\ref{fig:bifocal}~(a-f). First, Figs.~\ref{fig:bifocal}~(a,b) show the axial field cross-section, $p(x,y=0,z)$, using the pseudo-spectral simulation method and measured experimentally, respectively. We observe that the reconstructed field accurately matches the target foci, and the experimental results agree with the simulations. The corresponding transverse field distributions at $z=105$ mm are shown in Figs.~\ref{fig:bifocal}~(c,d) where sharp focusing is observed. The focal spots present larger dimensions in the axial $(z)$ direction than in transverse ones, as expected from limited-aperture holographic lenses, where the spatial spectral components in axial direction are limited by the finite-aperture source\cite{o1949theory}. Axial (measured at $x=25$ mm and $y=0$ mm) and transversal (measured at $z=105$ mm and $y=0$ mm) cross-sections are shown in Figs.~\ref{fig:bifocal}~(e,f), respectively. Excellent agreement is observed between simulation and experiment for the axial field profile at the focal region. A small secondary lobe located before the main one appears experimentally. The transverse profile shows a small lateral shift of $\pm$ 0.5 mm in both experimental foci towards the $x$-axis origin. 

Note that, due to diffraction, the geometrical focus of a geometrically focused source do not correspond to the acoustic focus of the source \cite{o1949theory}. In our case, the target location was set to $z=105.5$ mm, the acoustical focus of an equivalent focused source of same frequency and aperture in water peaks at $z=99.8$ mm and the focus of the lens peak at $z=100.4$ mm and $z=100.1$ mm in simulations and experiments including the skull phantom, respectively. These shifts corresponds to errors of 0.6\% and 0.3\%, respectively, showing the accuracy of the focusing performance of the holographic lenses.

\begin{figure*}[t!]
	\centering
	\includegraphics[width=1\linewidth]{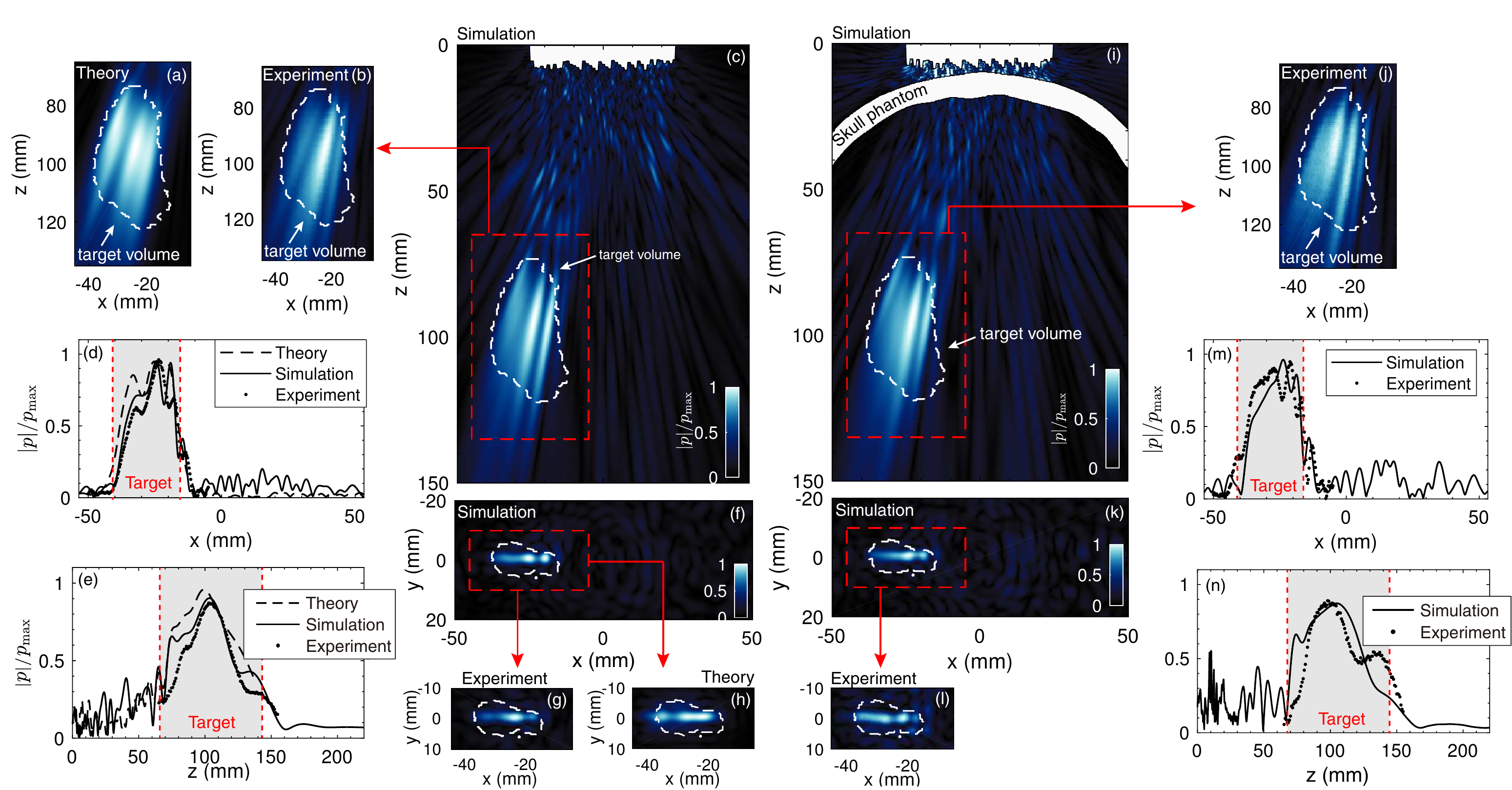}
	\caption{Volumetric hologram results. (a,b,c) Theoretical, experimental and simulated axial pressure distribution in water. (d,e) Transversal and axial field cross-sections in water. (f,g,h) Simulated, experimental and theoretical transversal field distribution in water. (i,j) Simulated and experimental axial pressure distribution including the skull phantom. (k,l) Corresponding transversal pressure distribution and (m,n) transversal and axial cross-section, respectively.}
	\label{fig:hologram}
\end{figure*}

\section{Self-bending beams}

The previous results show that holographic phase plates can retain phase information of multiple foci. Using this idea, we can set more complex targets following the shape of functional structures found in the CNS. Here, we set the target holographic pattern to a beam following a curved trajectory as those reported in homogeneous media without aberrating layers using active sources or metamaterials \cite{zhang2014generation,li2015metascreen,chen2018broadband}. As the aberration layers will be present in the real application, known analytical methods to calculate the phase of the 3D trajectory are, in principle, not available \cite{zhang2014generation}. Instead, we make use of a TR method: a set of virtual sources were placed along this trajectory and their back-propagated field were calculated. We set a factor of $(z/z_{max})\exp(iz k_z)$ to compensate for the amplitude and phase of each source to set the main direction of propagation, being $k_z$ the axial component of the wave-vector and $z_{max}$ the distance to the farthest virtual source (45 mm in this example). A sketch of the target trajectory is shown in Fig.~\ref{fig:selfbending}~(a). The axial and transversal cross-sections of the forward propagated field in water are shown in Figs.~\ref{fig:selfbending}~(a,b), respectively. We observe that using TR method the self-bending beams can be obtained, and the beam accurately follows the target trajectory. Using simulation and a lens made of elastic material a similar result is obtained, as shown in Figs.~\ref{fig:selfbending}~(c,d). The experimental tests show a similar pressure field distribution in comparison with theory using the Rayleigh-Sommerfeld integral and simulations using pseudo-spectral methods. A lateral shift of the peak pressure location of 0.3 mm in the $x$ direction is observed at $z=30$ mm and $y=0$ mm in the experiments. 

Finally, when the aberration layer of the skull phantom is included the corresponding holographic lens also reconstructs the target acoustic image with curved trajectory, as shown in Figs.~\ref{fig:selfbending}~(e,f). A similar lateral shift of the peak pressure location in the experiments, of 0.25 mm in the $x$ direction is observed at $z=30$ mm and $y=0$ mm. The measured pressure field inside the skull phantom agree the simulation. Note that the transversal size of the curved beam at $z=30$ mm is 1.11, 1.07 and 1.19 times the wavelength in water for the theoretical calculation, and for the simulations in water and including the skull phantom, respectively. Both results demonstrate that using TR methods self-bending beams following a target curve can be obtained inside the skull phantom using acoustic holographic lenses.

\section{Volumetric holograms overlapping CNS structures}
Going further, we designed a holographic lens which produces an acoustic image that fits the right human hippocampus volume. The holographic surface was placed near the occipital/parietal bones to adapt the acoustic image to the elongated geometry of the hippocampus. However, we locate the lens at the center of the skull symmetry plane in order to demonstrate the steering capabilities of this holographic lens. The lens generation process is based on the TR method with multiple virtual sources covering the target area (see Methods section for further details). Figure~\ref{fig:hologram} summarizes the results for both, water and including the aberration layer of the skull phantom.

On the one hand, Figs.~\ref{fig:hologram}~(a,b,c) show the forward-propagation field distribution of the holographic lens designed for water obtained using the theoretical, experimental and simulation results, respectively. First, we observe a good agreement between experiments, simulation and theory in both axial (Figs.~\ref{fig:hologram}~(a-c)) and transversal field distributions (Figs.~\ref{fig:hologram}~(f-h)). The beam is steered in the correct direction and a broad focal spot is generated. The transversal and axial field cross-sections are shown in Figs.~\ref{fig:hologram}~(d,e). The diffraction-limited image is reconstructed and the field is enhanced mainly at the target volume. To quantify the performance, we define the overlapping volume as the overlapping volumes of the target region and the region of the acoustic pressure field under a threshold corresponding to the half of the peak amplitude. In particular, using this lens we obtain in water an overlapping volume of 29.7\%, 20.1\% and 19.0\% for the theoretical calculation, simulation and experiment, respectively. 

On the other hand, the field distribution produced by acoustic holographic lenses including the skull phantom is shown in Figs.~\ref{fig:hologram}~(i-n). The experimental forward-propagation field distribution overlaps a similar volume in comparison with simulation result. Both holographic images present the same qualitative performance and provide a similar overall covering of the interest zone. In particular, an overlapping volume of of 21.1\% and 23.2\% was obtained in simulation and experiment, respectively. In addition, both axial (Fig.~\ref{fig:hologram}~(i,j)) and transversal (Figs.~\ref{fig:hologram}~(k,l)) field distributions are similar of those produced in water without the skull phantom, showing that, first, limited-diffraction holographic volumes can be reconstructed and, second, the aberrations produced by the skull phantom on these complex beams can be compensated at the source plane by the acoustic holographic lenses. Finally, the transversal and axial cross-sections, shown in Figs.\ref{fig:hologram}~(m,n), show that the experimental and simulated acoustic holographic lens produces a field enhancement that matches the target distribution. 

Note that the spatial bandwidth of the image is limited by the diffraction limit and the spatial bandwidth of the acoustic holographic lens \cite{melde2016holograms}. In this case, the holographic lens focuses at $z\approx 74.1 \lambda$ (100 mm), and its limited-aperture is only $a = 18.5\lambda$ (25 mm). Therefore, the transversal components of the wave-vector are band-limited and the performance of the holographic lens at this distance is restricted. Using lenses with larger aperture will improve the quality of the holographic acoustic image, and, therefore, the total overlapping volumes.

\begin{figure*}[t]
	\centering
	\includegraphics[width=1\linewidth]{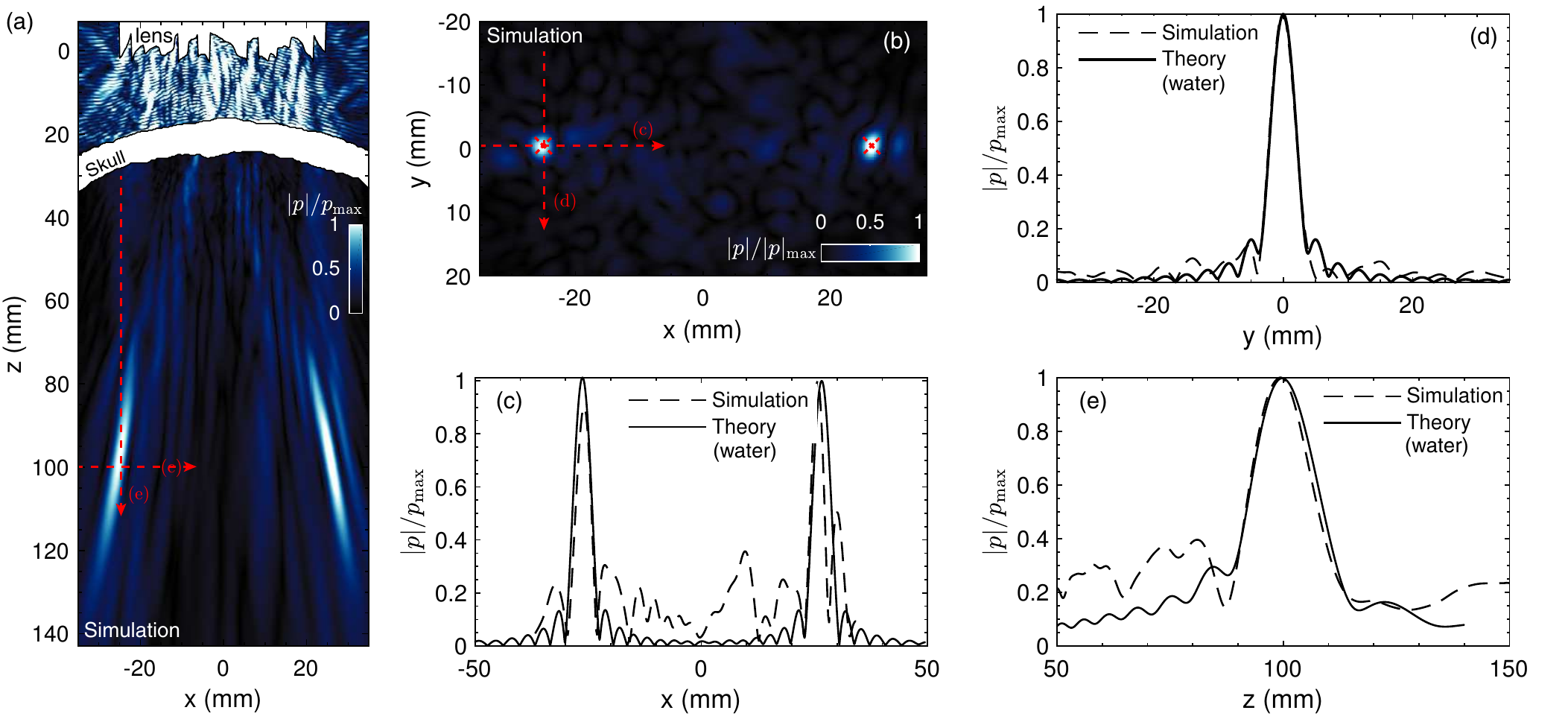}
	\caption{Simulation results for the bifocal holographic lens designed for a realistic skull. (a) Axial cross-section of the pressure distribution at $y=0$ mm. (b) Transversal cross-section at $z=100$ mm. (c) Lateral cross-section at $z=100$ mm and $y=0$ mm. (d) Lateral cross-section at $z=100$ mm and $x=-26=0$ mm. (e) Axial cross-section at $x=-26=0$ mm and $y=0$ mm. }
	\label{fig:real_bi}
\end{figure*}

\section{Hologram simulation using a realistic skull}
It is worth noting here that the impedance of the available 3D printing material used to manufacture the skull phantom is soft compared with the skull bone. In this way, the phase aberrations produced by a real skull will be stronger than the ones observed in the previous experiments. To demonstrate the focusing performance of the proposed lenses in a realistic situation a set of simulations were performed using the acoustical parameters of skull bones. {The parameters were derived using the same the CT data and were listed in Section \ref{sec:meth:skull}}. 

First, the results of the bifocal lens simulation using a realistic skull are summarized in Fig~\ref{fig:real_bi}. First, the sagittal cross-section of the absolute value of the pressure field at $y=0$ mm is shown in Fig.~\ref{fig:real_bi}~(a). We can see that the lens focuses at two clear spots, almost at the target distance. The corresponding traversal cross-section is shown in Fig.~\ref{fig:real_bi}~(b) measured at $z=100$ mm. In fact, good agreement is found between the simulations using a realistic skull and the calculations using the Rayleigh-Sommerfeld integral considering homogeneous water media. These two focal spots are generated together with small amplitude secondary lobes. {The amplitude of the side lobes is -8.86 dB below the peak pressure in the theory in water and -5.16 dB in the simulation including the skull.} To quantify the focusing performance of the lens, we show in Fig.~\ref{fig:real_bi}~(c) the transversal cross-section at $z=100$ mm and $y=0$ mm. The lateral shift of the left focus is -26.3 mm and -26.0 mm for the theoretical prediction and for the simulation, respectively. A relative error of 1.1\% was committed. These small lateral shifts are of the order of the experimental test shown previously with the 3D printed phantom. {Moreover, the amplitude of the side lobes in the simulation using a realistic skull are 0.3 times the peak pressure.} These side-lobes present higher amplitude in the lateral cross-section joining both foci (Fig~\ref{fig:real_bi}~(c)) than in the lateral cross-section measured at $x=0$ mm, as shown in Fig~\ref{fig:real_bi}~(d).  Finally, Fig.~\ref{fig:real_bi}~(e) shows the axial pressure distribution measured at the location of the right hippocampus. The axial peak location of the simulation including a realistic skull ($z = 99.3$ mm) matches the location of the corresponding peak pressure using the theory in water ($z=99.8$ mm). A relative error of 0.5\% is obtained, showing that the aberrations of a real skull can be mitigated using holographic lenses even when the target acoustic field presents a complex structure.

\begin{figure}[b]
	\centering
	\includegraphics[width=1\linewidth]{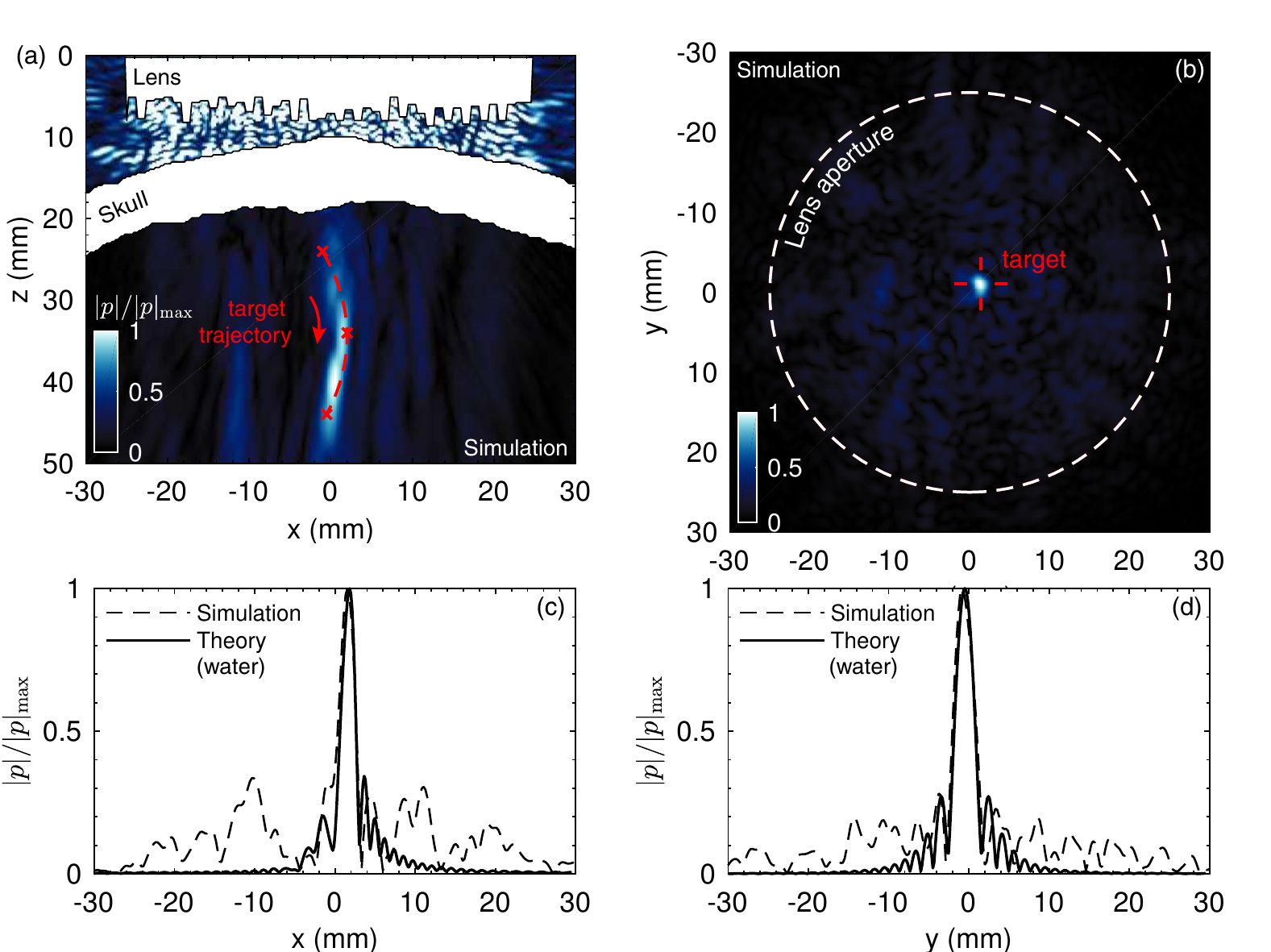}
	\caption{Simulation results for the self-bending holographic beam designed for a realistic skull. (a) Axial cross-section of the pressure distribution at $y=0$ mm. (b) Transversal cross-section at $z=35$ mm. (c) Lateral cross-section at $z=35$ mm and $y=0$ mm. (d) Lateral cross-section at $z=35$ mm and $x=2$ mm.}
	\label{fig:real_bend}
\end{figure}

\begin{figure*}[t]
	\centering
	\includegraphics[width=1\linewidth]{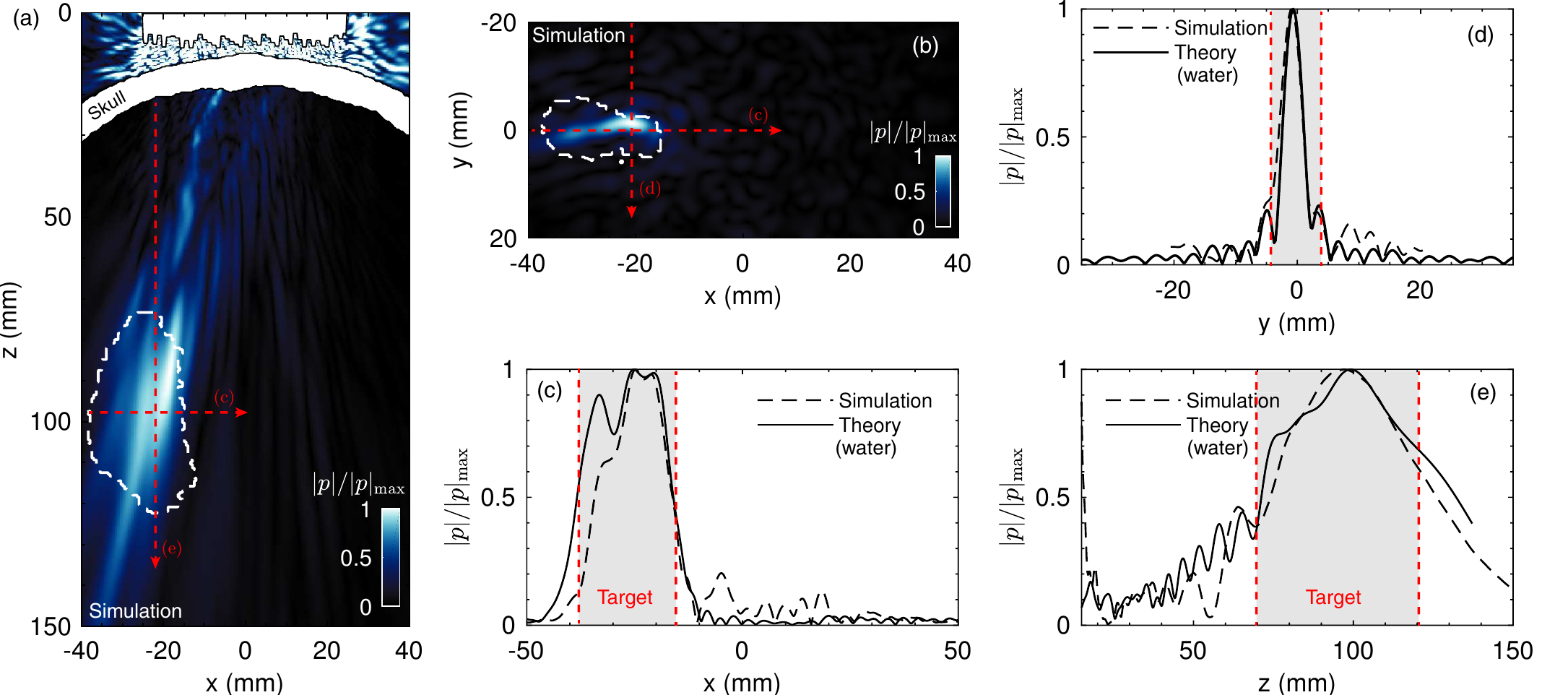}
	\caption{Simulation results for the volumetric hologram designed for a realistic skull. (a) Axial cross-section of the pressure distribution at $y=0$ mm. (b) Transversal cross-section at $z=95$ mm. (c) Lateral cross-section at $z=95$ mm and $y=0$ mm. (d) Lateral cross-section at $z=95$ mm and $x=-26=0$ mm. (e) Axial cross-section at $x=-26=0$ mm and $y=0$ mm. }
	\label{fig:real_vol}
\end{figure*}

Second, the results for the self-bending beam simulation inside a realistic skull are shown in Fig.~\ref{fig:real_bend}. The sagittal cross-section is shown in Fig.~\ref{fig:real_bend}~(a), measured at $y=0$ mm, together with the location of the target marked in red dashed line. In this case the performance of the lens to produce such a complex beam is reduced as compared with the previous cases, as can be seen by the presence of secondary lobes. This is mainly caused by the generation of strong stationary waves between the skull bone and the lens. However, the peak pressure follows the target trajectory and the location of the peak pressure matches the center of the curve. A clearer picture is given in Fig.~\ref{fig:real_bend}~(b), that shows the transversal field distribution measured at $z=35$ mm. Here, a sharp focal spot is visible and location of the peak pressure almost matches the location of the target focus. A lateral shift of $+0.22$ mm, that corresponds to one numerical grid step in the simulation was found in the $x$ direction. The corresponding transversal field distributions are shown in Fig.~\ref{fig:real_bend}~(c) for the $x$ direction and in Fig.~\ref{fig:real_bend}~(d) for the $y$ direction, respectively. Here, the reference calculations using theoretical methods in a homogeneous medium (water) are also shown for comparison. The location of the focal spots observed in both lateral directions for the simulations including the realistic skull are in excellent agreement with the corresponding focal spots in water. The width of the focal spot obtained using both calculation methods also is in agreement. The main discrepancy is the presence of secondary lobes in the simulated field, presenting a peak amplitude of 0.36 times the pressure at the focus. The amplitude of these lobes is 1.7 times larger in direction in which the beam is bent.

Third, a holographic lens was designed with the target of producing a volumetric hologram overlapping with the right hippocampus volume, and in this case including the acoustic properties of a realistic skull. The resulting forward-simulated pressure field is shown in Figs.~\ref{fig:real_vol}~(a-e). First, the sagittal cross-section of the magnitude of the pressure field at $y=0$ mm is shown in Fig.~\ref{fig:real_vol}~(a). The produced field focuses around the target volume, shown in dashed white lines. The beam is steered in the direction of the right hippocampus while the transducer axis remains normal to the skull surface. The transversal cross-section at $z=95$ mm is shown in Fig.~\ref{fig:real_vol}~(b). While the acoustic field is focused into the target volume, there exists areas not covered by the beam, mainly in the outermost regions away from the transducer source. To quantify the focusing performance, field cross-sections along the corresponding dashed lines are given in Figs.~\ref{fig:real_vol}~(c-e). The lateral cross-section at $z=95$ mm  and $y=0$ mm is shown in Fig.~\ref{fig:real_vol}~(c). As a comparison, we also plot the corresponding cross-section of a holographic lens designed to produce the same hologram in water using the theoretical Rayleigh-Sommerfeld integration. The simulated beam using the realist skull and the theoretical prediction in water mostly overlaps, being the energy of the beam concentrated into the target volume. Small side-lobes with an amplitude 5.3 times smaller than the peak pressure are observed in the simulations including the realistic skull. Note that the corresponding acoustic intensity of the side lobes is about 30 times smaller than the intensity at the focus. The volume of the beam, defined as the total volume of the beam under a threshold of 0.5 times the peak pressure, roughly overlaps with the target volume. The overlapping volume between the target and the volumetric acoustic hologram is 29.7\% for the theoretical calculation in water and 21.4\% for the simulation including the skull. The transversal cross-section along the $y$ axis is given in Fig.~\ref{fig:real_vol}~(d), where excellent agreement is found between the two configurations. Finally, the axial cross-section along the $z$ axis is shown in Fig.~\ref{fig:real_vol}~(e). In this case, the field presents remarkable side-lobes before and after the target region, but its amplitude is lower than half of the maximum pressure. Note this direction corresponds to the beam axis and the pressure distribution of the corresponding focused beam presents an elongated shape due to the limited aperture of the source. In this case, a good agreement is also found between the axial pressure distribution of the simulation including the realistic skull and the theoretical prediction in water.

\section{Conclusions}
{We have shown that using 3D printed acoustic holograms it is possible to conform diffraction-limited ultrasonic fields of arbitrary shape compensating the aberrations of a the human skull. In particular, experimental tests using a 3D printed skull phantom and numerical simulations using a realistic skull were performed to accurately generate multiple focal holograms, self-bending beams and volumetric holographic fields overlapping a target CNS structure. The proposed approach using holographic lenses represents a step forward when compared with the existing solutions using phase arrays, since it opens new venues to develop reliable and cost reduced ultrasonic applications.

The quality of the reconstructed acoustic images is related to the diffraction limit and the spatial bandwidth of the holographic lens, which depends on the spatial aperture of the lens, the number of pixels of the lens, and the frequency of the beam \cite{melde2016holograms}. In this work, we target a human hippocampus using a single holographic lens of only 50 mm aperture and operating at a frequency of 1.1 MHz. The reported experimental results inside a skull phantom are in good agreement with theory and simulations. Only small shifts, of the order of one wavelength (1.4 mm in water), were found between the target location and the field produced by the holographic lens. These shifts can be caused by experimental reasons that include small positioning error between the lens and the curved surface of the phantom and can be corrected using optimization methods during lens design. It is worth noting here that the phantom used in the experiments presents a smaller acoustic impedance than a real skull. However, full-wave simulations performed using the density, sound speed and attenuation values of skull-bone show that these arbitrary fields can also be produced in a realistic situation. The generated acoustic fields inside the skull were in good agreement with those produced in water. This shows that the aberrations produced by the skull can be mitigated by using holographic lenses even when a complex field is required. 

{Moreover, using the proposed methodology diffraction is captured exactly as compared with Fraunhofer or angular spectrum methods, leading to a better accuracy of the generated acoustic fields. In addition, the holographic lens design is based on resonating slabs, which include not only the refraction over a curved plate, but the resonating waves inside the lens.}

Phased-arrays are efficient but their high-cost can be prohibitive to spread out some of the incoming ultrasonic transcranial therapy treatments. Using phased arrays the ultrasonic beams can be adjusted in real time and monitored using MRI, obtaining a precise location of the acoustic focus into de CNS. Nevertheless, the number of elements of the phased-array systems can be insufficient to produce a complex volumetric ultrasonic field that matches a specific CNS structure. The use of holographic lenses present several advantages to produce complex volumetric patterns. First, the cost of a 3D printed lens is low as compared with phased-array systems. Second, each pixel in a holographic lens acts as an element of the phased-array with fixed phase.  Due to the high number of passive sources in a holographic lens, more than 4000 for the small lenses considered here, complex patterns can be generated. However, once the lens is designed its focal distance and the spatial features of the holographic image remains fixed and, in principle, it is not possible to steer the ultrasonic beam in real time with accuracy. For this reason, the technique is specially relevant for the treatment using a single sonification of structures or in the sonification of large volumes. 

The concept shown in this paper opens new doors to optimize and widespread incoming therapy treatments such as ultrasound-assisted blood-brain barrier opening for drug delivery and neuromodulation, or ultrasonic imaging of the central nervous system using low-cost devices. Considering the emergence of metamaterials and their huge flexibility, we also advance incoming biomedical applications of active holographic metasurfaces for the generation of complex fields in the central nervous system. }

\begin{acknowledgements}
This work was supported by the Spanish Ministry of Economy and Innovation (MINECO) through Project TEC2016-80976-R. NJ and SJ acknowledge financial support from Generalitat Valenciana through grants APOSTD/2017/042, ACIF/2017/045 and GV/2018/11. FC acknowledges financial support from Ag\`encia Valenciana de la Innovaci\'o through grant INNCON00/18/9 and European Regional Development Fund (IDIFEDER/2018/022).
\end{acknowledgements}

\appendix

	\section{Measurement setup}\label{appendix}

\begin{figure}[h]
		\centering
		\vspace*{2mm}
		\includegraphics[width=1\linewidth]{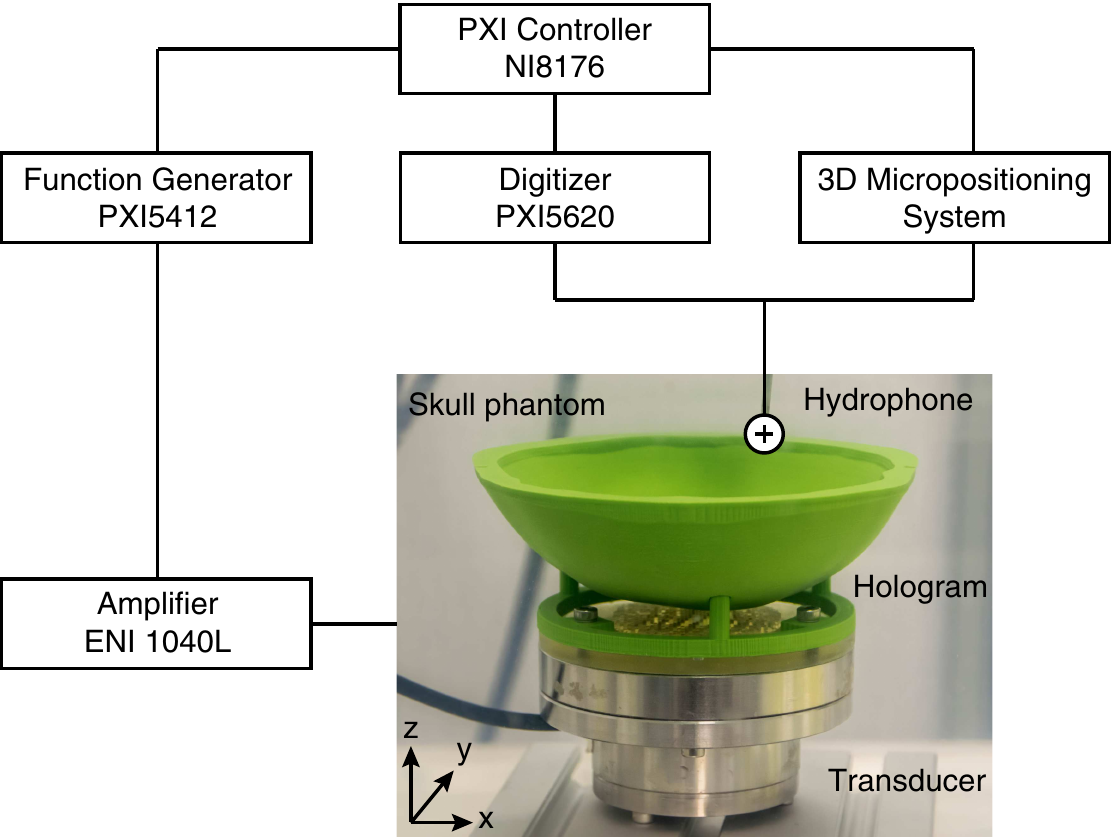}
		\caption{Experimental setup showing the block diagram and skull phantom inside the water tank with the ultrasonic source and the acoustic hologram at the bottom.}
		\label{fig:figplates}
	\end{figure}

	The experiments were conducted inside a $1\times0.75\times0,5$ m water tank filled with degassed and distilled water at 26$^\circ$. The ultrasonic transducer was composed by a single element circular piezoceramic crystal (PZT26, Ferroperm Piezoceramics, Denmark) mounted in a custom designed stainless-steel housing with aperture $2a = 50$ mm as shown in Fig.~\ref{fig:figplates}~(e). The transducer was driven with a 50 cycles sinusoidal pulse burst at a frequency of $f = 1.112$ MHz by a signal generator (14 bits, 100 MS/s, model PXI5412, National Instruments) and amplified by a linear RF amplifier (ENI 1040L, 400 W, 55 dB, ENI, Rochester, NY). The pressure field was measured by a needle hydrophone with a 500 $\mu$m active diameter (149.6 mV/MPa sensitivity at 1.112 MHz, Model HNR-500, Onda) calibrated from 1 MHz to 20 MHz. The source amplitude was set low enough to avoid any nonlinear effects in the propagation, we measure 1.8 kPa at the focus for the bifocal lens. The hydrophone signals were digitized at a sampling rate of 64 MHz by a digitizer (model PXI5620, National Instruments) averaged 100 times to increase the signal to noise ratio.

	An 3D micro-positioning system (OWIS GmbH) was used to move the hydrophone in three orthogonal directions with an accuracy of 10 $\mu$m. For the bifocal lens experiment the scanning an area for the sagittal cross-sections, $p(x,z)$, covers from -40 mm to 40 mm in the $x$ direction and from 82 mm to 143 mm in the $z$ direction, using a step of 0.5 mm in both directions, for the transversal cross-section planes, $p(x,y)$, covers from -40 to 40 mm in the $x$ direction and from -10 mm to 10 mm in the $y$ direction, using a step of 0.3 mm. For the self-bending lens, the scanned area covers from 20 to 50 mm in the $z$ direction, from -5 to 5 mm in the $x$ direction, and from -5 to 5 mm in the $y$ direction, using the same spatial steps. Finally, for the volumetric holograms the scanned area covers from 65 to 135 mm in the $z$ direction, from -45 to -5 mm in the $x$ direction, and from -10 to 10 mm in the $y$ direction, using the same spatial steps. All the signal generation and acquisition processes were based on a NI8176 National Instruments PXI-Technology controller, which also controlled the micro-positioning system. Temperature measurements were performed throughout the whole process to ensure no temperature changes of 0.5$^\circ$ C.



\end{document}